\documentclass[%
groupedaddress,
preprint,
 amsmath,amssymb,
 aps,
]{revtex4-1}
\usepackage[utf8]{inputenc} 
\usepackage{graphicx}
\usepackage{dcolumn}
\usepackage{bm}

\begin{document}

\title{Localized vortex beams in anisotropic Lieb lattices}

\author{Cristian Mejía-Cortés}
\email{ccmejia@googlemail.com}
\affiliation{Programa de Física, Facultad de Ciencias Básicas, 
Universidad del Atlántico, Puerto Colombia 081007, Colombia}
\author{Jorge Castillo-Barake}
\affiliation{Programa de Física, Facultad de Ciencias Básicas, 
Universidad del Atlántico, Puerto Colombia 081007, Colombia}
\author{Mario I. Molina}
\affiliation{Departamento de Física, Facultad de Ciencias, Universidad 
de Chile, Casilla 653, Santiago, Chile}

\begin{abstract}
    We address the issue of nonlinear modes in a two-dimensional waveguide
    array, spatially distributed in the Lieb lattice geometry, and modeled by a
    saturable nonlinear Schr\"odinger equation. In particular, we analyze the
    existence and stability of vortex-type solutions finding localized patterns
    with symmetric and asymmetric profiles, ranging from topological charge
    $S=1$ to $S=3$. By taking into account the presence of anisotropy, which is
    inherent to experimental realization of waveguide arrays, we identify
    different stability behaviors according to their topological charge. Our
    findings might give insight on experimental feasibility to observe these
    kind of vortex states.
\end{abstract}

\maketitle

\section{Introduction}\label{intro} 

Localized vortex beams (VBs) are interesting objects that can emerged in nonlinear
optical lattices. They are characterized by feature several bright spots, which
are spatially distributed according to the refractive index distribution,
maintaining a nontrivial phase distribution in space. This phase circulates
around a singular point, or central core, changing by $2\pi S$ times in each
closed loop around it (with $S$ being an integer number). The integer number $S$
is known as the topological charge (TC) of the vortex. In general, optical
vortices have been envisioned as a mean to codify information using their TC
value in classical~\cite{PhysRevLett.88.013601} and quantum~\cite{Mair:2001aa}
regimes. A stable vortex is capable of delivering angular momentum to a nearby
object, hence, one of its most remarkable applications are optical tweezers in
biophotonics, where they are useful due to their ability to affect the motion of
living cells, virus, and molecules~\cite{Zhuang188,Favre-Bulle:2019aa}. Other
applications can be found in optical systems
communication~\cite{Barreiro:2008aa} and spintronics~\cite{PhysRevLett.86.4358}.

Given an optical nonlinear medium of a specific geometry, we are interested in
ascertaining whether vortex solitons can exist in principle and its stability
properties. For homogeneous media, a self-focusing Kerr nonlinearity leads to
the collapse of any two-dimensional solution~\cite{PhysRevLett.15.1005}. 
Bright solitons can be stabilized by adding a transversal periodic refractive 
index, but, any non fundamental solution becomes unstable as they collapse upon
propagation. However, recently has been shown that by inserting spatially
alternating gain and losses, stable localized vortices can exist under the same
nonlinear scenery~\cite{Kartashov:16}.  On the other hand, photorefractive
materials, where the linear refractive index can be modulated externally by
light, vortex solitons have been envisioned
theoretically~\cite{PhysRevLett.93.063901} and observed
experimentally~\cite{PhysRevLett.92.123903} in two-dimensional square lattices
at focussing regime. The nonlinear response for these crystals features a
saturable nature and its limit value depends of the voltage applied to the
crystal, among other parameters.  Hence, photorefractive crystal becomes a
terrific workbench to observe localized patterns. However, its underlying
mechanism (the photorefractive effect) is intrinsically anisotropic, which for
example causes deformation of the induced waveguides~\cite{Armijo:14}.  

During the last decade, femtosecond laser inscription technique has shown solid
and versatile results in fabrication of optical waveguides, both in
amorphous~\cite{Szameit_2010} and crystalline dielectric
materials~\cite{Castillo:s}. In the latter case, it would be desirable to
achieve nonlinear photonic lattices by direct laser writing in photorefractive
materias, e. g., barium titanate (BaTiO3), lithium niobate (LiNbO3), strontium
barium niobate (SBN), etc. In this fashion, it could be possible to tailor an
unique experimental setting where a saturable nonlinearity can be combined with
a strong refractive index variation, transversal to propagation axis of light.
Motivated for this scenery, we aim at delving into the stability properties of
such vortex solitons as supported modes of a Lieb lattice, in the presence of
anisotropy, from a tight-binding approach. The choice of the Lieb lattice is
interesting since it is one of several 2D lattices that display a flat band in
its spectra, from which originates compacton-like
states~\cite{PhysRevLett.114.245503, PhysRevLett.114.245504}.  In that respect,
our results could give some quantitative insight, in terms of parameters of the
system, about experimental observation of these kind excitations. Hence, we hope
that our findings stimulate experimental realization of such photonic lattices.

\section{Model}\label{sec:model}

We start from a model of coupled optical beams in a nonlinear periodic medium.
From Maxwell's equations one obtains a paraxial equation which  is further
approximated by a coupled-modes expansion, ending up in the discrete nonlinear
Schr\"odinger (DNLS) equation. This model, in its general form, can be
written as 
\begin{equation} -i \frac{dU_{\vec{n}}}{dz} =\sum_{\vec{m}}
C_{\vec{n},\vec{m}}U_{\vec{n}} +F(|U_{\vec{n}}|^{2})U_{\vec{n}},
\label{modelo2d} \end{equation}
where $U_{\vec{n}}$ is the electric filed amplitude at waveguide $\vec{n}$, $z$
is the coordinate along the longitudinal direction, $C_{\vec{n},\vec{m}}$
are the coupling terms between sites $\vec{n}$ and $\vec{m}$. The last term
assume the role of nonlinear response of the system, that in our case will
be modeled by the real function
$F(|U_{\vec{n}}|^{2})=-\gamma/(1+|U_{\vec{n}}|^{2})$, standing
for a photorefractive media whose nonlinear coefficient $\gamma$ is proportional 
to the external applied voltage.

\begin{figure}[t]
    \begin{center}
  	\includegraphics[width=0.90\linewidth]{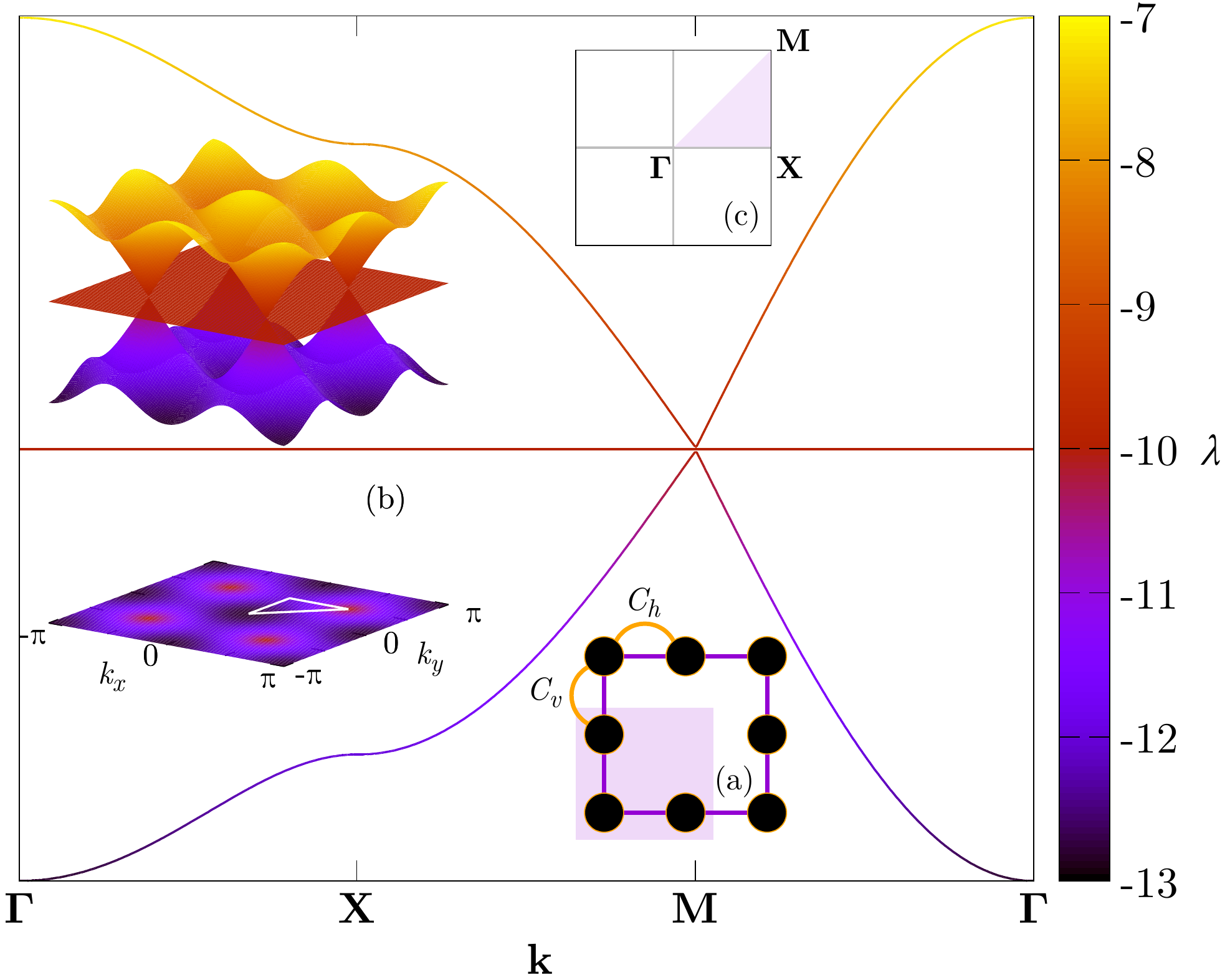}
    \end{center}
    \caption{Dispersion relation in the Lieb lattice for linear part of
    Eq.~(\ref{modelo2d}) along high symmetry points. Unitary cell of the lattice
    is represented with shadow region in (a).  Irreducible
    Brillouin zone, corresponding to this unitary cell, is delimited with a white
    triangle in (b) and sketched in (c).}
    \label{f1}
\end{figure}

We will use the Lieb lattice, a kind of a diluted square lattice, that
corresponds to a 2D version of the perovskyte structure [cf. Fig.~\ref{f1} (a)],
that has three sites in its unit cell (denoted by the shadow region), and a
lattice constant with length equal to $a$.  We will consider coupling to
nearest-neighbors only and we introduce anisotropy by setting the horizontal
coupling $C_h$ different from the vertical one, $C_{v}$.  By keeping the first
term in expansion of nonlinear response, i. e.,  $F(|U_{\vec{n}}|^{2})\approx
-\gamma$, and using the Bloch functions $U_{n,m}=u_{n,m}
\exp{[i(k_{x}n+k_{y}m)a+i \lambda z]}$ as solutions of the Eq.~(\ref{modelo2d}),
the dispersion relation for the Lieb lattice in the coupled mode approach
becomes
\begin{equation*}
   \lambda = -\gamma,\quad \lambda_{\pm} = \pm 2 \sqrt{C_{h}^2 \cos^{2}(k_{x}a) + 
   C_{v}^2 \cos^{2}(k_{y}a)}-\gamma.
\end{equation*}
Hence, we can observe that there exist two conical interaction bands and one
flat band, shown in Fig.~\ref{f1} (b). These bands meet at single points,
${\bf M}=(\pm k_{x}na,\pm k_{y}ma)$, called Dirac Points.

For the nonlinear case, we look for stationary solutions of
model~(\ref{modelo2d}) of the usual form, $u_{n,m}(z)=\phi_{n,m}\exp{(i\lambda
z)}$, where $\phi_{n,m}$ is the field amplitude which defines a complex spatial
profile of the solution, and $\lambda$ is the propagation constant. In this case
we have a nonlinear eigenvalue problem, given by
\begin{equation}
   \lambda u_{\vec{n}} = \sum_{\vec{m}} C_{\vec{n},\vec{m}} u_{\vec{n}} - 
   \gamma \frac{u_{\vec{n}}}{1+|u_{\vec{n}}|^{2}},
   \label{eq2}
\end{equation}
i.~e., a set of coupled nonlinear complex algebraic equations.  We are
interested in solutions that have an integer number ($S$) of $2\pi$ phase
changes in the azimuthal direction, so that, the phase profile twists in a
helical manner as the beam propagates. In such a case, the self-localized
solution is called a discrete vortex soliton, which have been predicted, both in
conservative~\cite{PhysRevE.64.026601} and dissipative~\cite{PhysRevA.83.043837}
system.

\section{Localized vortex type solutions}\label{modes}

As a particular setting for our analysis, we assume a focusing media by choosing
a positive value for nonlinear coefficient ($\gamma=10$).  In order to find
nonlinear VBs we proceed to solve Eq.~(\ref{eq2}) by implementing numerically a
Newton-type algorithm. By choosing proper and reasonable seeds, in each case,
and setting $C_v=C_h=1$ , we unveil families of stationary solutions with
different values of TCs. In addition, we perform a standard linear stability
analysis (see appendix in ref~\cite{PhysRevA.86.023834}) for the whole set of
solutions.  From now on we use solid (dashed) lines to denote family regions
where VBs are stable (unstable).

\begin{figure}[t]
    \begin{center}
      \includegraphics[width=0.90\linewidth]{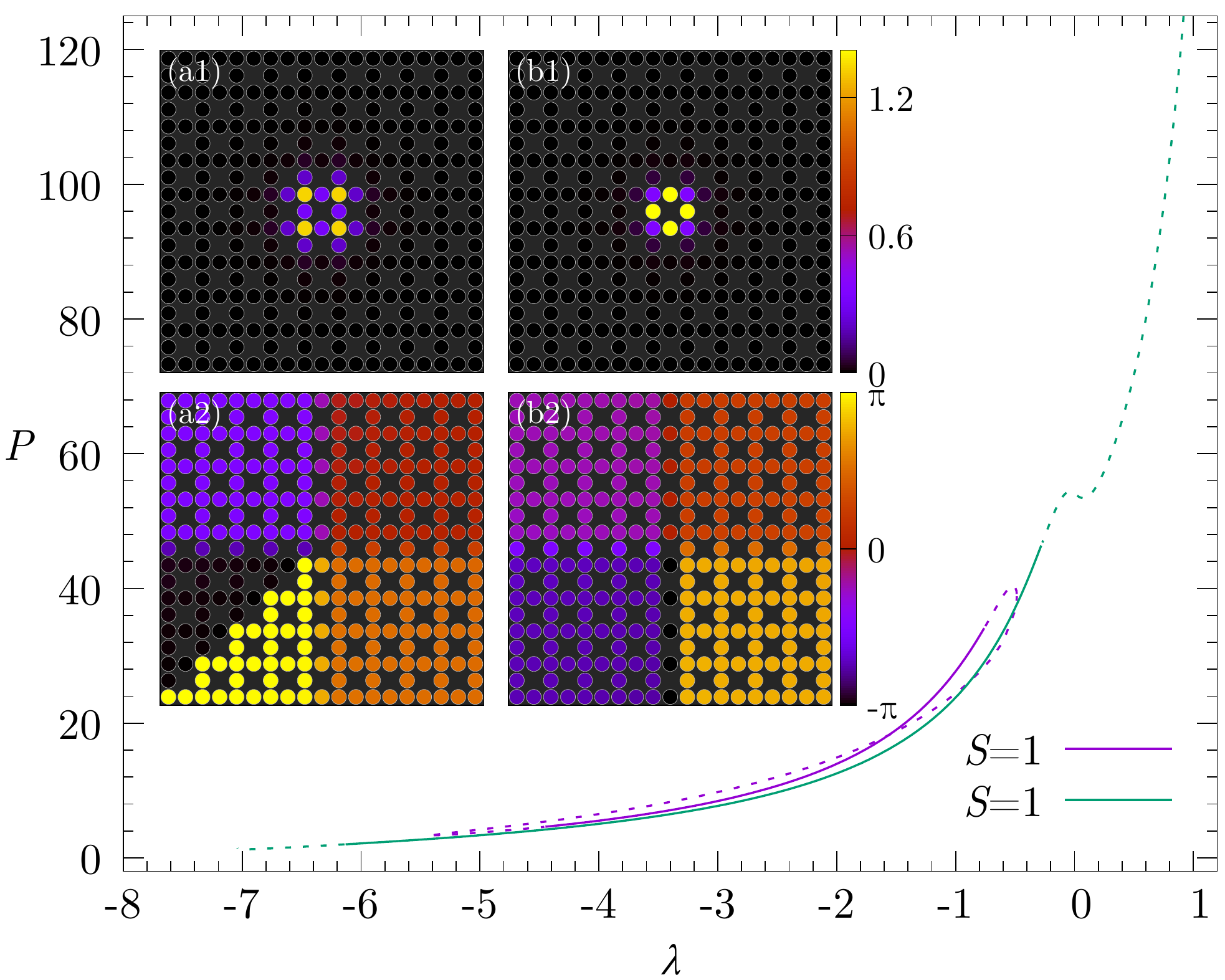}
    \end{center}
    \caption{$P$ vs $\lambda$ diagrams for two families of discrete vortex
    solitons with $S=1$. Amplitude (a1, b1) and phase (a2, b2) profiles, at
    $\lambda=-3.0$, correspond to solutions belonging to the families
    represented by the green and magenta curve, respectively.}
    \label{f2}
\end{figure}

We begin by presenting results of stationary vortex solitons with one
TC ($S=1$). Figure~\ref{f2} shows the power content $P~=~\sum_{\vec{n}}
|u_{\vec{n}}|^{2}$ vs $\lambda$ diagrams corresponding with two
different kind of localized modes. In both cases, their main spots are unfolded
along an off-site place circuit: a square one for solutions belonging
to the family represented by a green curve, and a ring one for those 
corresponding with magenta curve. Amplitude and phase spatial profiles, 
for both modes, are sketched in Fig.~\ref{f2} (a1, a2) and Fig.~\ref{f2} (b1,
b2), respectively.

It is worth mentioning that modes belonging to the  green curve exist for almost
the entire $\lambda$ domain: high power solutions are not restricted, and
low-power ones exist for values close to the maximum value of the upper linear
band, $\lambda=2\sqrt{2}-\gamma$, at ${\bf \Gamma}=(0,0)$ point
[cf.~Fig.~\ref{f1}].  One reason for this behavior could be that spatial profile
for this kind of mode can be seen as a superposition of four fundamental bright
solitons, which bifurcate from the fundamental band-edge
mode~\cite{PhysRevA.98.053845}.  However, unlike the fundamental ones, here the
phase distribution is not trivial. The four main spots are locked in a special
manner, so, the solution does not bifurcate from the linear band and the green
curve turns back at $\lambda\approx -7.04$. On the contrary, for the magenta
curve, the lowest power value solution corresponds to $\lambda\approx -5.38$
where the family turns back increasing again its power up to $\lambda\approx
-0.52$. From here, power decreases and the family turns back again. In general,
both types of solutions destabilize for low and high power values (at least in
our finite range of parameters), nevertheless, they span sizable stable regions.
It is interesting to point out that there exist two different types of ring mode
solutions with the same power at equal values of propagation constant
($\lambda\approx -1.55$), one stable and the other one unstable.

\begin{figure}[t]
   \begin{center}
      \includegraphics[width=0.90\linewidth]{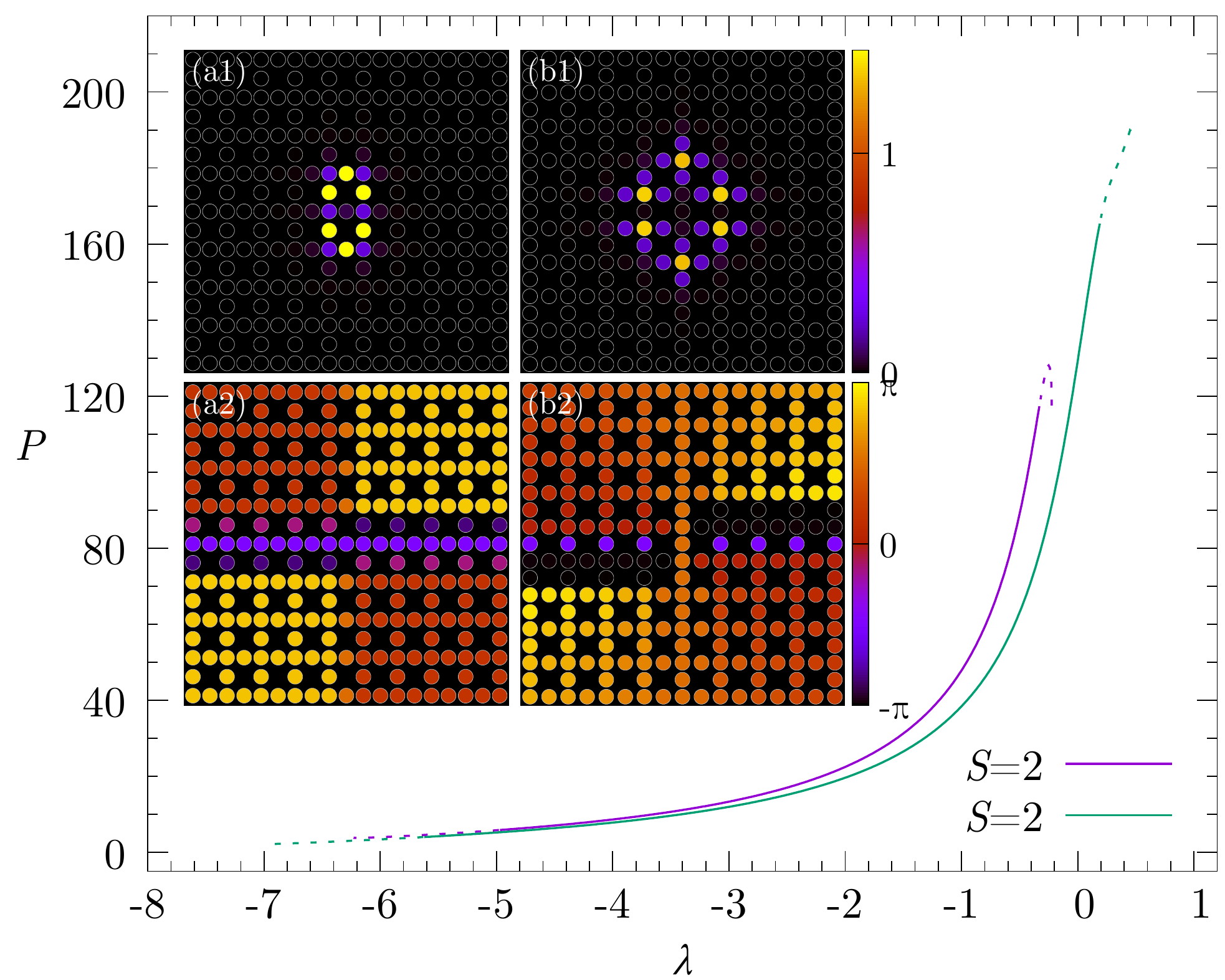}
	\end{center}
    \caption{$P$ vs $\lambda$ diagrams for two families of discrete vortex
    solitons with $S=2$. Amplitude (a1, b1) and phase (a2, b2) profiles, at
    $\lambda=-3.0$, correspond to solutions belonging to the families
    represented by the green and magenta curve, respectively.}\label{f3}
\end{figure}

We focus now on the results of stationary vortex solitons with two TCs ($S=2$).
Figure~\ref{f3} shows $P$ vs $\lambda$ diagrams corresponding with another pair
of localized modes. For the present case, their main spots are unfolded along
irregular hexagonal paths: a small one for solutions belonging to the family
represented by magenta curve, and a big one for those corresponding with green
curve. Amplitude and phase spatial profiles, for both modes, are sketched in
Fig.~\ref{f3} (a1, a2) and Fig.~\ref{f3} (b1, b2), respectively.  It is
important to point out that amplitude profiles for last modes do not resemble
the $S=1$ ones discussed at first, who had a smaller size.  Instead, it should
be emphasized that both of them look like asymmetric localized patterns
vertically orientated.  

The small hexagonal mode possesses two vertical edges with length equal to $2a$
and four diagonal edges with length equal to $\sqrt{2}a$ [Fig.~\ref{f3} (a1)].
This kind of pattern can be presumed as the juxtaposition, along the vertical
axes, of two ring modes [cf. Fig.~\ref{f2}(b1,b2)] properly oriented. The last
one possesses two vertical edges with length also equal to $2a$ and four
diagonal edges with length equal to $\sqrt{8}a$ [Fig.~\ref{f3} (b1)]. On the
other hand, the spatial phase profile for small hexagon mode [Fig.~\ref{f3}
(a2)] displays abrupt changes along closed paths around the centroid of the
pattern.  This behavior is not exclusive for periodical
systems~\cite{Soto-Crespo:09}.  Even more, its variation does not occur
continuously. Perhaps, that information is missing because the DNLS model does
not account for any light propagating outside of the waveguides as evanescent
waves. By contrast, the phase profile for the second mode [Fig.~\ref{f3} (b2)]
exhibits a smoothly variation. When power goes down the family represented by
the green curve gets closer to the linear band and it turns back at
$\lambda=-6.97$.  The magenta curve behaves similarly, but it does not come so
close to this band and its returning point is located at $\lambda=-6.23$. 

\begin{figure}[t]
    \begin{center}
        \includegraphics[width=0.90\linewidth]{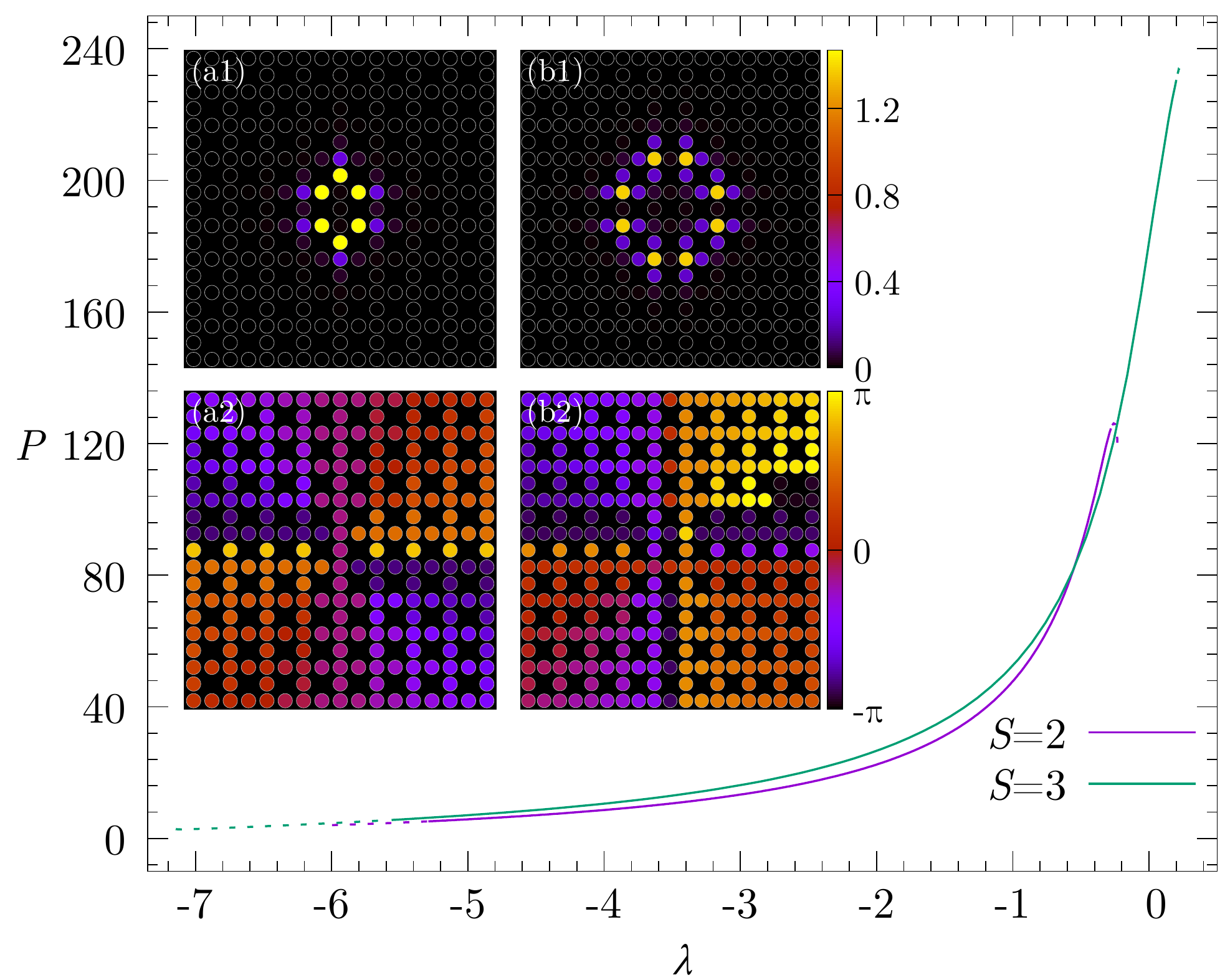}
    \end{center}
    \caption{$P$ vs $\lambda$ diagrams for two families of discrete vortex
    solitons with $S=2$ (magenta) and $S=3$ (green). Amplitude (a1, b1) and
    phase (a2, b2) profiles, at $\lambda=-3.0$, correspond to solutions
    belonging to these families represented by the green and magenta curve,
    respectively.}
    \label{f4}
\end{figure}

Finally, we examine couple more cases of stationary vortex solitons with two and
three TCs ($S=2$ and $S=3$), respectively. Figure~\ref{f4} shows $P$ vs
$\lambda$ diagrams corresponding with these two localized modes. In both cases,
their main spots are unfolded along closed path trajectories: a hexagon one
equivalent to that sketched at Fig.~\ref{f3}(a1,a2), for solutions belonging to
the family represented by a green curve, and an octagon one for those being part
of green curve. Amplitude and phase spatial profiles, for both modes, are
sketched in Fig.~\ref{f4} (a1, a2) and Fig.~\ref{f4} (b1, b2), respectively.
Despite hexagonal amplitude profiles depicted in Fig.~\ref{f3}(a1) and
Fig.~\ref{f4}(a1) looks very similar, even with the same segment lengths, the
number of lattice sites enclosed by those spatial trajectories are not the same.
In this case, the juxtaposition would between two on-site ring modes.
Nevertheless, although the TC for both cases is the same, the spatial phase
profile for the present case exhibits a smoothly variation. On the other side,
modes with octagonal profile possesses two vertical and two horizontal edges
with length equal to $2a$ and four diagonal edges with length equal to
$\sqrt{8}a$ [Fig.~\ref{f4} (b1)]. Here again, phase profile also varies softly
displaying clearly three TCs.

In general, we note that stable solutions with higher values of TCs, and a
reasonably well-behaved phase variation, display a bigger spatial profile. Our
results show that for vortex solitons with $S=1$ their pattern are composed of
four main spots, two less than those with $S=2$ (hexagon). For the $S=3$ case,
their amplitude profile exhibit eight main spots (octagon) two more than
previous one.  Apparently, as many peaks are present in the amplitude profile
the solutions can endow a higher value ot TC. This approach looks like effective
to find VBs with TC greater than $S=3$. However, and after following
this insight, we could not find solutions with this attribute.  It
seems that the highest allowed value for the TC is restricted by the order of
discrete point-symmetry group of the lattice~\cite{PhysRevLett.95.123902}. 

\section{Stability against anisotropy}\label{modes}

\begin{figure}[h]
    \begin{center}
        \includegraphics[width=0.9\linewidth]{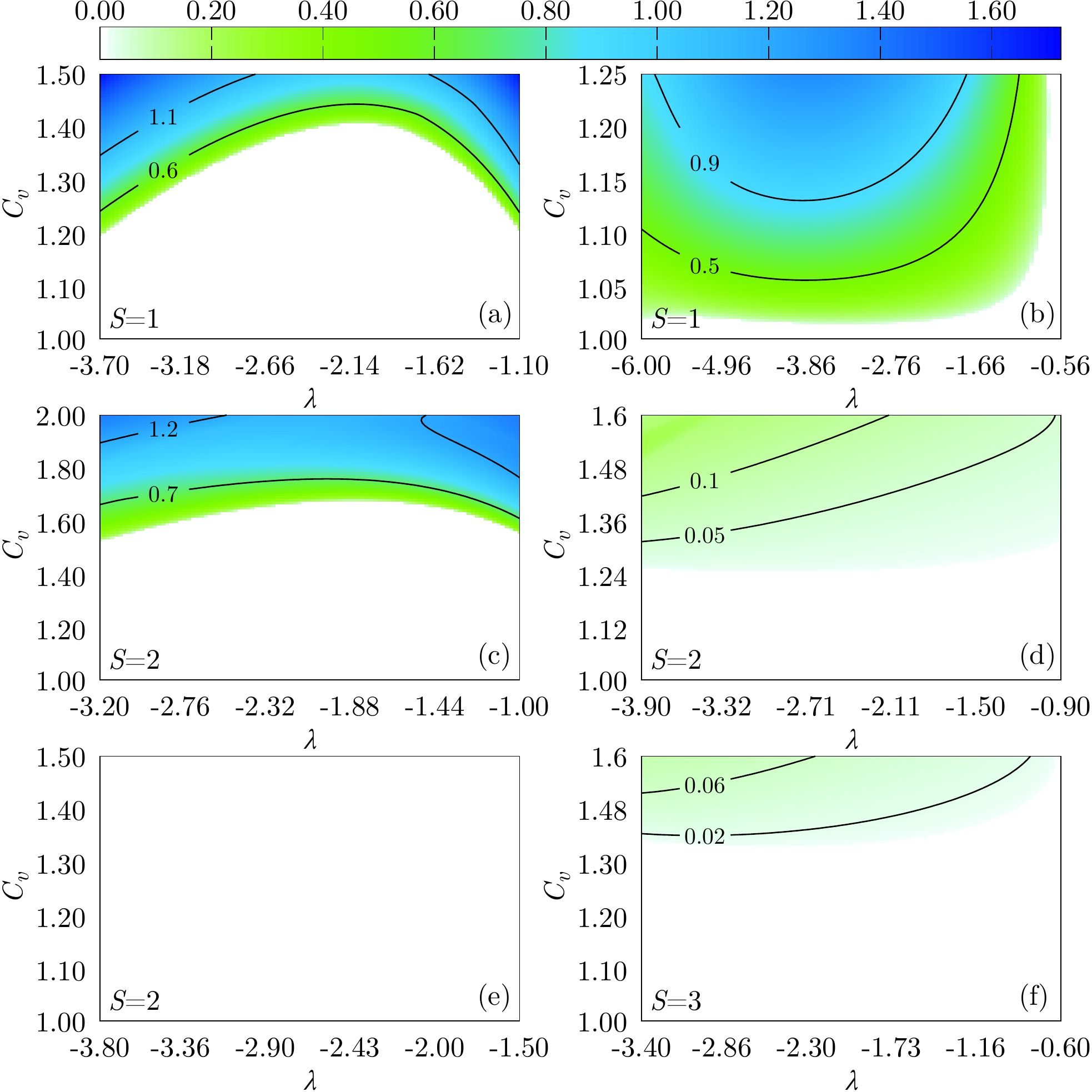}
    \end{center}
    \caption{Colormap plots for stability domains as function of $\lambda$ and
    $C_v$ for diverse vortex solitons. Modes endowed with one TC: (a) ring
    amplitude profiles and (b) square amplitude profiles. Modes endowed with two
    TC with three different (c)-(e) asymmetric hexagonal amplitude profiles.
    Mode with (f) octagonal profile and three TCs.}
    \label{f5}
\end{figure}

So far, we have analyzed the existence and stability of localized nonlinear
vortex modes with different TCs in terms of $P$ vs $\lambda$ diagrams. Having in
mind the anisotropic constraint, imposed by the photorefractive
effect~\cite{Terhalle:2007aa}, in a hypothetical experimental realization of
these kind of lattices, we extended our analysis to the anisotropic case ($C_v
\neq C_h$). By adding this quantity as a new parameter in our workspace, we
could figure out how reasonable would be the experimental observation of this
kind of localized patterns.  To do that, we must first unveil the existence
domain, for one specific type of localized vortex solution, as a function of the
$\lambda$ and $C_v$ parameters. That is, we must make sure that the whole set of
solutions in $(\lambda, C_v)$ domain feature the same characteristics, namely,
TC and spatial amplitude distribution.  In general, we have observed that
anisotropy shrinks $P$ vs $\lambda$ diagrams, both in existence and stability.
Therefore, starting from the isotropic case, and by increasing adiabatically the
strength of anisotropy, we explore a region of parameters to identify cartesian
boundaries of a $(\lambda, C_v)$ subspace.  Once this has been outlined, we
perform a detailed scan of its inside.  Figure~\ref{f5} displays the stability
diagrams for vortex modes reported above in terms of the anisotropy of the
system. The size of $(\lambda, C_v)$ subspace is
$\{\lambda_1,...,\lambda_N\}\times\{C_{v_1},...,C_{v_N}\}$ being $N=100$, i.~e.,
each one of these plots are sampled with $10^{4}$ points, where each point
corresponds to one specific stationary solution. 

For ring modes with $S=1$ we can appreciate from Fig.~\ref{f5}(a) a considerable
inner region where this kind of solution remains stable. Depending of the
$\lambda$ value, a higher strength of anisotropy can be supported for these kind
of vortices, finding that maximum value allowed here for anisotropy is
$C_v\simeq 1.40$ at $\lambda\simeq -2.14$. A really different behavior occurs
when we analyze the stability properties of square modes with $S=1$.  From
Fig.~\ref{f5}(b) we note that the stable region is almost imperceptible; once we
move along the anisotropy axis, the solutions becomes unstable except for
$\lambda\simeq -0.56$, where a narrow stability window raises up to a strength
anisotropy value of $C_v\simeq 1.25$.
We turn now to stability regions for modes with two TCs ($S=2$) and asymmetric
hexagonal profiles. In the first case, for those modes displayed at
Fig.~\ref{f3}(a1,a2), we observe that strength of anisotropy allowed for the
off-site hexagon [Fig.~\ref{f5}(c)] is $C_v\simeq 1.66$ corresponding to
$\lambda\simeq -1.88$.  Next case, for those modes displayed at
Fig.~\ref{f3}(b1,b2), they also possesses a stable region but smaller than the
previous case with maximum value for anisotropy $C_v\simeq 1.34$ at
$\lambda\simeq -0.9$. In the last case, pertaining to the modes depicted at
Fig.~\ref{f4}(a1,a2), even when the existence domain is smaller than for the
first case their stable regions are very similar.
Finally, we scrutinize the stability region for modes endowed with three TCs
($S=3$), and an octagonal amplitude profile as those represented in
Fig.~\ref{f4}(b1,b2). Here, we also observe [cf. Fig.~\ref{f5}] a sizeable
stable region where stable vortex can support anisotropic values up to $C_v=1.6$ at
$\lambda\simeq -0.68$.

\section{Conclusions} 

Recapitulating, we have analyzed the existence and stability of VBs with diverse
values of TCs, in periodic media with saturable nonlinearity, as function of
parameters of the system. Families of stationary solutions with one, two and
three topological charges has been unveiled. In general, as we enforce our model
to obtain vortices with TCs greater than $S=1$, the number of main spots
increases, what leads to spatial amplitude profiles become larger. On the other
hand, by studying the role played in the system by the presence of anisotropy,
we identify domains of stability for each type of mode reported here. We showed
that those vortices that could be regarded as generated from two ring modes,
display larger stable regions than others.

We consider that model employed along this work might fit pretty well for
waveguide arrays imprinted in photorefractive crystals, therefore, we hope
that our findings stimulate fabrication for these photonic lattices, and
additionally they pave the way for the observation of these VBs.

\noindent {\bf Acknowledgments}: Powered@NLHPC: This research was partially
supported by the supercomputing infrastructure of the NLHPC (ECM-02). M. I. M.
acknowledges support from Fondecyt grant 1160177.

\noindent {\bf Disclosures}: The authors declare no conflicts of interest.


\bibliography{sample}


\end{document}